# Emerging Technology and Policy Co-Design Considerations for the Safe and Transparent Use of Small Unmanned Aerial Systems

Ritwik Gupta, Alexander Bayen, Sarah Rohrschneider, Adrienne Fulk, Andrew Reddie, Sanjit A. Seshia, Shankar Sastry, Janet Napolitano

December 2022

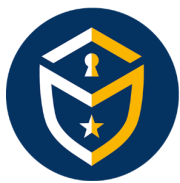

# Introduction

Over the course of the last decade, we have seen rapid technological progress in the space of small unmanned aerial systems (sUAS) which has enabled entirely new modes of operations for many organizations across industry, academia, and government. With disruption in industries such as logistics[1,2], defense[3], agriculture[4], and more, sUAS are steadily becoming commonplace in American society. Consequently, policy surrounding the safe and transparent use of sUAS has become an increased focus for agencies around the United States.

sUAS technology is developing at an exponential rate that policy cannot keep up with. Technology and policy co-design is critical to ensure sUAS are brought into everyday usage with minimal threat to society. In this work, we analyze the technology and policy design considerations by splitting the sUAS operations vertical into discrete elements: understanding how human operators interact with sUAS and other sUAS operators, how signals are sent securely to and between sUAS, how air traffic is coordinated in this dynamic new world, and how trust is built into onboard autonomy. These elements neatly organize the sUAS ecosystem into the analog, digital, and biological components that work together during sUAS operations. Further, since technology is in a constant state of upgrade, we examine how policymakers anticipate and respond to the constant emerging technologies.

To this end, the UC Berkeley Center for Security in Politics hosted a two-day workshop focused on the analysis of gaps in technology and policy development for sUAS. Composed of a group of over 60 experts in engineering and public policy, each slice of the sUAS vertical was thoroughly analyzed in individual sessions for best practices and open challenges that are critical for the successful integration of sUAS into society. These verticals include (1) human operations, (2) communications protocols, (3) air traffic control, and (4) autonomy.

---

[1] Amazon Unveils Smaller Delivery Drone That Can Fly in Rain. Spencer Soper and Matt Day. November 10, 2022. https://www.bloomberg.com/news/articles/2022-11-10/amazon-unveils-new-smaller-drone-for-faster-small-package-delivery

[2] Zipline. https://www.flyzipline.com/

[3] Defense Innovation Unit – Blue UAS. https://www.diu.mil/blue-uas

[4] The application of small unmanned aerial systems for precision agriculture: a review. Zhang and Kovacs. 2012. https://link.springer.com/article/10.1007/s11119-012-9274-5.

## 1. Responsible and Transparent Human Operations

Currently, human operators manage all aspects of sUAS operations; they decide when and where how to fly, how many sUAS to operate, and how these devices should interact with people and infrastructure around them. We want humans operating sUAS to be responsible and transparent in their actions and intentions. However, defining what responsibility is as well as how transparent to be are at the crux of this issue. Defining and detailing what these operations should look like are critical and will represent the base upon which new technology and policy should be constructed.

Critically, we identified three pain points where further technology and policy development are needed: training and certification, privacy policy, and inter-jurisdictional coordination.

*1.1 — Training and Certification*

Purchasing and operating sUAS has become exceedingly simple over the years. It is trivial to go online, purchase a suitable sUAS for recreational or commercial use, and begin operating within a day of delivery. An average citizen is able to acquire and operate a sophisticated sensor platform with complex flight dynamics with minimal training beyond what the sUAS's user manual provides. With an additional 30 minutes of optional "training", any sUAS hobbyist can begin flying their system with no additional safeguards in place.

Aerial operations are complex and covered by existing regulations. 30 minutes of training via The Recreational UAS Safety Test are insufficient to give a meaningful, lasting understanding of these regulations. The Federal Aviation Administration defines the Federal Aviation Regulations[5] which provide exhaustive statutes for airspaces, operating rules, licensure procedures, and much more. However, the average citizen operating sUAS is unaware of these regulations. Furthermore, enforcement of the FAR is lacking due to the sheer proliferation and relatively innocuous operation of these devices around the United States.

As an additional complicating factor, modern sUAS, including many hobbyist drones[6], offer the capability to fly autonomously, to a limited degree. sUAS are able to navigate and operate between pre-defined waypoints seamlessly. In this mixed operation mode where human and computer input are overlapping, conflicting inputs can result in unstable flight dynamics, potentially leading to a crash.

---

[5] Title 14 of the Code of Federal Regulations. https://www.ecfr.gov/current/title-14
[6] To a limited degree such as automated take off and landing, as well as automated "return home" features.

The FAA provides means by which sUAS operators can be licensed or certified to operate these devices. The Recreational UAS Safety Test (TRUST)[7] and the Remote Pilot Certificate[8] are mandated for recreational and commercial sUAS pilots, respectively. These certifications ensure that remote operators are aware of common airspace rules and regulations and are able to safely operate sUAS over people and near critical infrastructure. However, these certifications have no clear correlation with the quality of training provided to remote pilots in preparation for their certification exams. TRUST can be passed in ~20 minutes online with no practical sUAS operations experience with unlimited retakes allowed in any period of time with a fixed set of questions. As of May 21, 2022, the FAA had 538,172 recreational sUAS registered and only 257,437 TRUST completion certificates awarded[9] (~48%). The vast majority of recreational sUAS are unregistered[10]; the FAA's recreational sUAS process is under-utilized and under-prepares pilots for proper suAS operations.

Additionally, the material on the TRUST exam does not cover rapidly emerging sectors of sUAS technology such as autonomous operation, first-person view operation, and more. The pace of technological innovation is far outpacing the policies and regulations in place to safely introduce these technologies to modern society.

**Challenge:** The vast majority of private sUAS users do not require any form of licensing to operate and fly sUAS devices. Additionally, current training mechanisms fail to ensure operators are properly trained. Additionally, current certification processes are outdated and presume that sUAS are like traditional aircraft in many instances.

---

[7] The Recreational UAS Safety Test (TRUST). Federal Aviation Administration. July 25, 2022. https://www.faa.gov/uas/recreational_flyers/knowledge_test_updates

[8] Become a Drone Pilot. Federal Aviation Administration. August 16, 2022. https://www.faa.gov/uas/commercial_operators/become_a_drone_pilot

[9] Drones by the Numbers. Federal Aviation Administration. May 31, 2022. https://www.faa.gov/uas/resources/by_the_numbers/

[10] U.S. agency requires drones to list ID number on exterior. David Shepardson. February 12, 2019. https://www.reuters.com/article/us-usa-drones/u-s-agency-requires-drones-to-list-id-number-on-exterior-idUSKCN1Q12O9

**Opportunity:**  The FAA has been handed a tough charter. They must ensure the safety of all platforms that operate in the airspace while simultaneously not stifling new innovations. The rapid proliferation of sUAS technology took these choices out of the FAA's hands, resulting in the current fragmented training and licensing realm today. By working with sUAS manufacturers, the FAA still has the opportunity to standardize and rapidly roll out mandatory training that every recreational sUAS pilot must pass before operating their sUAS. Eventually, by forming an independent advisory board composed of members from industry, academia, and the legislature, the FAA might create a process by which pilot certification programs rapidly adapt to the introduction of new technologies to the sUAS market.

Many recreational sUAS activities are bespoke. There is a long tail of flight patterns and activities that cannot be covered by every single standardized training course. By staying ahead of the commercial sector via this proposed independent advisory board, the FAA can implement a "choose your own path" training course. Recreational pilots can choose activities they plan on engaging in with their sUAS, and their sUAS platform will automatically tailor a training course for them in accordance to FAA regulations.

*1.2 — Privacy*

The right to privacy is recognized in the United States by the Supreme Court[11] and generally protects against the intrusion of others into one's private affairs. Various jurisdictions have extended the right to privacy to the digital domain, providing individuals with remedy, or at the least, knowledge, about digital information collected about them or their activities. Remote surveillance of individuals without their knowledge or consent via aerial surveillance platforms has been found to be illegal[12]. However, the proliferation of sUAS platforms has democratized remote aerial surveillance outside of the hands of governments to the common citizen.

sUAS provide a means for powerful sensors, such as cameras, LIDAR sensors, and microphones, to encroach upon private land to record and store information which can directly violate other individuals' right to privacy. While prior case law guarantees some rights to airspace above their property for property owners[13], there is still much debate and conflicting legal theory about the rights to overflight and the legality of sUAS encroachment into private spaces.

---

[11] [Griswold v. Connecticut](#)
[12] [Leaders of a Beautiful Struggle v. Baltimore Police Department](#)
[13] [United States v. Causby](#)

**Challenge:** The technologies enabling the violation of an individual's right to privacy have proliferated widely into the consumer market. Legal theory and regulations have not reached an agreement as to what is an appropriate solution to this problem.

**Opportunity:** The current regulatory environment around sUAS privacy is handled on a case-by-case basis as new violations of privacy are enabled by new technologies in this sector. Lawmakers and the FAA have the opportunity to be proactive about the right to privacy with sUAS by building around the scaffold of existing remote surveillance laws. A two-headed approach of (1) self-regulation via better training procedures and (2) meaningful extensions of surveillance laws from the 2D ground plane to the 3D aerial plane can result in a landmark privacy bill protecting the rights of citizens before this problem becomes too widespread.

## 2. Safe Protocols and Policies for Communications and Cryptography

sUAS operations represent a form of communication. Messages need to be sent to a platform in order to operate it, messages need to be received from a platform containing feedback and data, and communications need to be in place between the pilot, the sUAS, and the surrounding infrastructure in order to fly safely. Protecting these communications channels and ensuring that communications are of high quality are therefore paramount to promoting healthy sUAS adoption.

*2.1 — Inter-pilot Communication Mechanisms*

With an uptick in recreational and commercial sUAS operating in increasingly congested environments, enabling effective communication between remote pilots has become a priority for technologists and policy makers alike. sUAS pilots are currently unable to communicate with other sUAS around them, nor are they able to communicate with infrastructure supporting the airspace they operate in. The ability to communicate, when needed, is critical for safety in aerial operations[14].

The concerns here are multiple: there are limited policies in place with what communications responsibilities remote pilots have with respect to each other and the airspace, technologies that allow for easy and simple communications regarding sUAS operation are nearly non-existent, and secure communications between operators and their sUAS are still a fledgling area for technology and policy.

Remote ID[15] rules mandate that most sUAS operating in the US broadcast their identity and their location, but no true two-way communication channel functionality has been defined by the FAA or other regulatory agencies. This gap in rule-making has arisen due to conflicts in schools of thought of how to treat the rapidly growing sUAS sector. On one side of the argument, requiring two-way communications from all sUAS activity greatly increases the burden that pilots face when managing their radios, if they possess radios in the first place. Additionally, increasing communications result in an increasingly congested radio spectrum. Conversely, an increase in two-way communications can result in avoided air traffic accidents.

---

[14] Improving Pilot/ATC Voice Communication in General Aviation. Daniel G. Morrow and O. Veronika Prinzo. Office of Aviation Medicine. July 1999.
[15] UAS Remote Identification. Federal Aviation Administration. November 10, 2022. https://www.faa.gov/uas/getting_started/remote_id

Combined with advances in peer-to-peer communications networks and swarm operation modes for sUAS, there are multiple, simple means by which instant communications channels can be established between multiple sUAS operators. This functionality has corollaries to proximity chat in popular video games and virtual meeting platforms, in which multiple users form many-to-many communications networks as they approach each other in large virtual environments.

**Challenge:** sUAS platforms do not provide a standardized ability to enable communication with other sUAS pilots, and limited policies exist regarding the responsibilities of remote pilots to communicate with other pilots or infrastructure around them.

**Opportunity:** sUAS manufacturers can develop mesh communications technologies and protocols that enable pilots to share information with each other on an as-needed basis. This can reduce the cognitive burden associated with a completely open two-way communications system. The commercial sector has the opportunity to collaborate in order to create a flexible and open standard that all sUAS platforms can integrate into.

After a period of time, regulators should convene a panel of experts that will review the state of the world for two-way communications for sUAS traffic to provide recommendations and formal standardization of an accepted protocol/framework. This activity will solidify industry actions while providing feedback that will guide the further maturation of such a protocol.

2.2 — *Secure Protocols and Dedicated Frequency*

Since sUAS platforms necessitate extensive wireless digital communication for operations commands as well as data transfer and storage, the reliability and security of these channels is of great importance. A series of dropped command packets can lead to unstable flight dynamics, whereas malicious interception and decryption of information can lead to the loss of potentially sensitive data collected by the sUAS platform.

We live in a world where wireless spectrum availability is increasingly congested. Congestion can lead to interference, resulting in decreased reliability of communications between a source and receiver. This may result in aviation accidents in the form of aerial collisions or unplanned sUAS landings or crashes. Ensuring the reliability of control signals is paramount to robust and safe sUAS operation. While there are efforts such as DARPA Spectrum Collaboration Challenge[16] that attempted to create methods which enable the seamless sharing of wireless frequencies, dedicated frequency bands may serve as a better alternative to protecting this increasingly fragile ecosystem of sUAS communications. However, with frequencies high in demand and low in supply, the task of prioritizing which use cases receive protected allocation is an unenviable one.

Additionally, wireless communications themselves may not be secured in any way. Any person who is able to pick up on wireless transmissions (mostly everyone with a sufficient antenna between the source and receiver) is able to intercept these communications and understand them with minimal additional effort. With videos, telemetry, and other sensor information available for capture at any point, encryption of network traffic is essential to protect against malicious interception and interference.

Unfortunately, the regulations surrounding encrypted communications traffic are unclear and conflicting at times. Modern WiFi setups are encrypted by default, but amateur radio transmissions, such as the ones used by first-person-view sUAS platforms, may not be allowed to "scramble" their transmissions[17].

**Challenge:** In an increasingly contested wireless frequency domain, preventing benign or malicious interference or interception of signals is an increasingly difficult problem. Technology enables all portions of sUAS communications to be encrypted, but regulation does not mandate encryption, nor does it provide a clear path for those who would like to encrypt their communications.

**Opportunity:** The FAA and FCC can converge on a set of rules and technologies that result in a decrease in spectrum congestion. By guiding the growing commercial sector on the proper use of technologies such as spectrum hopping, encryption, and collaborative spectrum sharing, regulators can stay ahead of spectrum congestion.

---

[16] Spectrum Collaboration Challenge. DARPA. https://www.darpa.mil/program/spectrum-collaboration-challenge
[17] 47 CFR § 97.113 - Prohibited transmissions. https://www.law.cornell.edu/cfr/text/47/97.113

By collaborating with experts from academia and industry, the government (to include agencies such as the FAA, FCC, and NIST) can define a qualitative framework under which certain spectrum congestion mitigations must be applied and the desired outcomes from these measures that should be met.

The FCC, on a biannual basis, should review these policies and set penalties for sUAS manufacturers who do not comply with spectrum decongestion and communications rules. On the other side of the coin, the FAA should define penalties for sUAS pilots who knowingly skirt communications rules.

# 3. Protecting and Evolving the Air Traffic Control System

The air traffic control system has existed federally since mid-1936. Created to handle the meteoric increase in fixed-wing airplane travel, air traffic control plays a crucial role in ensuring the safe and efficient operation of air travel. It is responsible for directing the movement of aircraft, avoiding collisions, and providing information and support to pilots. Air traffic control has successfully adapted to the introduction of new modes of flight and technologies over the decades such as an increase in recreational flights, autopilot systems, and advanced runway operations mechanisms. Critically, these technologies were adopted over the span of many years, giving the Federal Aviation Administration and the air traffic system ample time to adapt to these changes.

The rapid introduction of sUAS to airspaces globally has led to the air traffic control system to be stretched thin. sUAS technology is evolving rapidly, leading to a rapid aging of the processes codified by the air traffic control system. Existing technologies required to monitor the rise in traffic and systems in the air are overwhelmed, and new technologies to monitor all air traffic are years out from introduction. Malicious, or otherwise detrimental, flight patterns are becoming increasingly common.

In this section, we will discuss the measures that must be taken to ensure the ongoing success of the air traffic control system. We will discuss the importance of security measures, incorporation of new technologies, and adaptability. By taking these steps, we can ensure that the air traffic control system continues to serve its crucial role in the safe and efficient operation of air travel.

*3.1 — Integration with Traditional Air Traffic Management*

The problem of mixed autonomy in air traffic management (ATM) is one of the key problems for which no architectural solution has emerged as a clear future paradigm. As a result, the corresponding jurisdictional division of authority and subsequent policies to support it has not been developed. In many settings, e.g. urban flight, there is no concept of operations for mixed autonomy, which would integrate some form of shared traffic control (human, on the ground, for classic aviation, with human on the ground for (semi)-automated aircraft mixed with manned traffic). While ideas of delegation of authority by the FAA to local/state governments have emerged, there is currently no framework for it. When operating over the built environment, there is the question of the category of airspace (would it be general airspace - FAA, or other). This would also probably require the re-design of low altitude airspace to incorporate (1) mixed autonomy, and (2) contingency management (in particular for unmanned aircraft). Contingency management is easier over water, train tracks and specific dedicated emergency landing areas, which would be an integral part of the airspace redesign. There is also a large number of regional / local airfields that could be used and would benefit from the rebirth of some activity linked to these new aircraft.

Both at Federal and State level, in order to push the development of such a framework (institutional, jurisdictional, policy), there would need to be a specific value proposition embraced by elected officials. Such a value proposition exists for very specific verticals, for example organ/medical sample transport (there is a competitive impetus to do so immediately—currently air-operated by commuter helicopter companies in the US, and drones in some countries in Europe). As the market expands, a redesign of the airspace to support such operations must occur together. Realistically, there needs to be a sufficiently large market built before there is pressure to redesign. This is also slowed down by the emergence of the e-VTOL paradigm (which still needs a few years before reaching a level of economic maturity to open new markets).

Concepts of operations need to be developed that make the space below 400ft airspace viable technologically and economically. Examples of such markets exist—with various vertical requirements—in the New York geography where air travel into Manhattan and into major surrounding airports has shown to be economically viable and compatible with noise and emission requirements. In ways similar to various degrees of autonomy on the ground, e.g. "self-driving" cars, specific levels of autonomy will lead to more viable concepts of operations, especially closer to cities.

For these new paradigms, new and existing technologies need to provide the proper tools, technologies, and procedures for anomaly detection for contingency management. These safety considerations are part of a larger set of security considerations so the safeguards developed through operational considerations cannot be circumvented by hostile attacks. The process of integration of drone operations (or flights with a certain degree of automation) might thus be a gradual process like on the road. For specific use cases, autonomous vehicles are certified on the road - for example slow shuttles in semi-secluded corridors. The same might happen as a way to gradually push the envelope of integration of manned and unmanned aircraft, and with it traffic management.

**Challenge:** Delegation between the various levels of government authorities for mixed autonomy is still a developing story, and enough market pressure to get these policies implemented is not there yet. Additionally, technologies and policies to define contingency plans in the case of failures—critical for ATM—do not exist.

**Opportunity:** The academic and industry sectors must collaborate to develop a cohesive plan for how sUAS technologies will handle, and critically, communicate failures and anomalies during flight operations. These technologies and recommendations will put sufficient pressure on policy-making bodies to codify the best of these into standard practice, ensuring that we can rapidly reach a point at which mixed autonomy is integrated into traditional ATM.

*3.2 — Countering Malicious sUAS Behaviors*

sUAS platforms have been used to conduct localized warfare abroad over the last decade. With increased sUAS proliferation and their availability to anyone without any sort of background check, the threat of domestic attacks via drones is not only increased, but a realized threat that has been processed by local judicial systems across the country[18].

Technologies to counter sUAS platforms are nascent and have lacked serious investment outside of the military sector. With counter sUAS (cUAS) solutions developed primarily for the military market[19,20], the needs and constraints of the domestic market are largely unaddressed. With a non-trivial portion of domestic sUAS flights occurring over buildings, infrastructure, and people (albeit against regulations), countering such sUAS behavior presents a risk to human populations below. In prior civil aviation literature, the International Civil Aviation Organization explicitly discourages governments from shooting down civilian aircraft[21], instead opting to intercept with other manned aircraft and guiding the target to a safe location. These guidelines are written with the assumption that all flight is manned; this is not true of unmanned sUAS operations.

---

[18] A Drone Tried to Disrupt the Power Grid. It Won't Be the Last. Brian Barrett. November 5, 2021. https://www.wired.com/story/drone-attack-power-substation-threat/
[19] Counter Unmanned Aerial Systems. Northrop Grumman. https://www.northropgrumman.com/what-we-do/land/counter-unmanned-aerial-systems-c-uas/
[20] CounterUAS. Anduril. https://www.anduril.com/capability/counter-uas/
[21] Manual concerning interception of civil aircraft: consolidation of current ICAO provisions and recommendations. International Civil Aviation Organization. 1984. https://digitallibrary.un.org/record/877?ln=en

Regulations regarding when, how, and where cUAS technologies are applicable are largely non-existent. With sUAS platforms being operated by US persons on US soil, difficult legal considerations at local, state, and federal levels arise regarding the rights of individuals or the government to counter malicious (or perceived to be malicious) sUAS behavior.

**Challenge:** Technologies to identify, track, and counter malicious sUAS behavior have been fielded in military contexts. Overall, R&D in this area is limited due to a lack of investment from agencies concerned with domestic security. Civilian law enforcement agencies have no impetus to invest in such technologies due to a lack of regulation and law regarding their authority to counter malicious sUAS behavior.

**Opportunity:** Malicious use of sUAS is a problem that has to be handled by appropriate law enforcement agencies. The FBI and DHS should work with the FAA to define when sUAS activity is malicious, what remedies law enforcement have when there is a known threat from sUAS, and when and how cUAS platforms may be used. A concrete doctrine will inform lawmakers to pass these authorities into federal and state laws.

## 4. Ensuring Trustable and Knowable Autonomy

sUAS platforms are steadily becoming more autonomous. The entire controls, navigation, perception, and to a degree, communications stack have a degree of automated intervention built in. Spurious human input that may destabilize the sUAS is automatically filtered out, and algorithms perform object classification and detection, planning, control and other tasks from the multitude of sensor inputs available on the sUAS platform. With this trend of moving from automation to autonomy, the task of ensuring that all autonomy on board is trustworthy, auditable, and reliable comes to the forefront of sUAS-related concerns.

*4.1 — Trustworthiness and Robustness*

Almost every system on a commercial or civil aircraft is mandated to have multiple forms of failover or redundancy in the case of system losses. Software on aircraft, with few exceptions[22], follows rigorous secure programming guidelines with real-time constraints and strict audit procedures. Modern artificial intelligence (AI), on the other hand, seems to run counter to this culture of robustness and interpretability that is built into modern aerospace software engineering. Notoriously functioning as a black box, deep neural networks and other forms of AI can be uninterpretable, unauditable, and lack guarantees of safe operation in uncertain and stochastic environments. AI system explainability, verification and validation, and safety engineering are growing fields of research and implementation which concern themselves with ensuring that AI systems can provide some assurances about their operational boundaries, but work is still nascent.[23]

With AI systems increasingly present on sUAS platforms, there is a rapidly increasing gap between what is technically possible and what is properly regulated. Autonomous sUAS flights are common in rural areas overseas, with entire missions carried out with no human intervention. Failures of these AI-enabled controls systems can be disastrous in urban environments where there is a high chance of human casualties.

---

[22] Boeing Charged with 737 Max Fraud Conspiracy and Agrees to Pay over $2.5 Billion. Department of Justice. January 7, 2021.
https://www.justice.gov/opa/pr/boeing-charged-737-max-fraud-conspiracy-and-agrees-pay-over-25-billion

[23] Toward Verified Artificial Intelligence. Seshia, Sadigh, and Sastry. Communications of the ACM, 65(7):46–55, 2022.
https://cacm.acm.org/magazines/2022/7/262079-toward-verified-artificial-intelligence/fulltext

Efforts have been made in the field of AI to simplify these models to those that have inherent interpretability or audit properties. For example, knowledge distillation[24] is the process of taking a black-box neural network and reducing it to a decision tree which has interpretable rules about the task that it is trained to carry out. Such mechanisms can enable complex capabilities to run on sUAS platforms while still providing a degree of reliability, interpretability, and robustness.

**Challenge:** Modern AI systems, to a large degree, are black box systems that are unable to be audited or interpreted. An increase in autonomous sUAS operations requires all onboard autonomy to have a degree of safety and reliability which is made difficult due to this very black box nature of AI systems.

**Opportunity:** Academia and industry must rapidly innovate in this area to come up with easily usable AI safety techniques particularly for the sUAS sector and the workloads present therein. Furthermore, the FAA, NTSB, and NIST, combined with experts from universities and industry, must come to an agreement on standardized engineering practices and benchmarks that onboard autonomy systems are required to comply with. Failure modes of these AI systems must have required failover systems, and emergency human takeover must be mandated at all times.

*4.2 — Attacks Against AI Systems*

AI models are known to have vulnerabilities[25] that let adversaries control the outcome of a model's decision process. These vulnerabilities span the entire lifecycle of an AI model, all the way from data poisoning attacks to model inversion. More often than not, vulnerabilities in AI models are not understood. Current AI engineering best practices notably do not suggest processes to monitor for these vulnerabilities, neither are methods to discover AI vulnerabilities easy to use, discover, or rectify. Recent progress from DARPA through its GARD[26] and Assured Autonomy programs[27] make some headway towards this goal, but a long road remains in order to truly guarantee the safety of AI systems.

---

[24] Distilling the Knowledge in a Neural Network. Hinton, Vinyals, and Dean. March 9, 2015. https://arxiv.org/abs/1503.02531
[25] Adversarial Machine Learning. Huang et. al. October 2011. https://archive.ischool.berkeley.edu/tygar/papers/SML2/Adversarial_AISEC.pdf

sUAS, especially those tasked with human interaction such as in delivery, construction, and security applications, are particularly ripe targets for AI attacks. Unexpected, unexplained, and indefensible behavior in the autonomy stack of sUAS platforms can result in human casualties or loss to critical infrastructure.

**Challenge:** AI models are vulnerable to a new class of cyber attacks that can be remotely triggered by adversaries. These vulnerabilities are poorly understood and modern AI engineering processes do not account for them during deployment.

**Opportunities:** There is a large gap in usability research into AI defense toolkits and robustness research. National research funding agencies such as the NSF must invest additional resources into intersectional areas of work such as AI applied to the sUAS space. Clear recommendations as to what classes of attacks sUAS autonomy must be robust against need to be put into place by NIST and the FAA, and an ongoing study as to what classes of attacks are feasible and which vulnerabilities will evolve to be pressing threats over time.

---

[26] DARPA GARD - Holistic Evaluation of Adversarial Defenses. https://www.gardproject.org/

[27] DARPA Assured Autonomy. https://www.darpa.mil/program/assured-autonomy

## Conclusion

The rapid technological growth observed in the sUAS sector over the past decade has been unprecedented and has left gaps in policies and regulations to adequately provide for a safe and trusted environment in which to operate these devices. The Center for Security in Politics at UC Berkeley, via a two-day workshop, analyzed these gaps by addressing the entire sUAS vertical. From human factors to autonomy, we recommend a series of steps that can be taken by partners in the academic, commercial, and government sectors to reduce policy gaps introduced in the wake of the growth of the sUAS industry.

# Acknowledgements

This work was supported by DARPA via the project Symbiotic Design of Policy and Technology for Emerging Technologies, a supplement to the DARPA LOGiCS project supported by the Symbiotic Design of Cyber-Physical Systems program grant FA8750-20-C-0156. The authors would like to thank Kaylie Washnock, Missy Cummings, Jay Merkle, Philip Mattson, Nicki Bugni, Naira Hovakimyan, Karl Johansson, Jarrod Knowlden, Raja Sengupta, Casey Hehr, George Pappas, Jay Meil, David Aguilar, Parimal Kopardekar, and Claire Tomlin for their contributions to this paper. This work would not have been possible without the dedicated effort of Kaylie Washnock, Jeanne Lynch, and Noah Kroloff at GSIS.

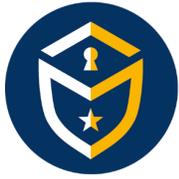